\documentclass{article}
\usepackage[margin=1.3in]{geometry}
\usepackage{authblk}
\usepackage{braket}
\usepackage{amsmath}
\usepackage{amssymb}
\usepackage{graphicx}
\usepackage{float}
\usepackage{fullpage}
\usepackage{hyperref}
\usepackage{titlesec}
\usepackage{multicol}
\usepackage{CJKutf8}

\hypersetup{
    colorlinks=true,
    linkcolor=blue,
    urlcolor=blue,
    citecolor=blue
}

\titleformat{\section}
  {\normalfont\large\bfseries}  
  {\thesection}{1em}{}  

\title{First Use of a Polarized $^3$He Neutron Spin Filter on the Back-n White Neutron Source of CSNS} 

\author[2,3]{Mofan Zhang (张墨凡)}
\author[2]{Zhou Yang (杨洲)}
\author[1,2,4]{Junpei Zhang (张俊佩)}
\author[1,2,4]{Chuyi Huang (黄楚怡)}
\author[1,2,4]{Tianhao Wang (王天昊)\thanks{Correspondence: 1 Zhongziyuan Road, Dalang, Dongguan, Guangdong 523000, China. Email: wangtianhao@ihep.ac.cn}} 
\author[1,2]{Yonghao Chen (陈永浩)}
\author[1,2,5]{Ruirui Fan (樊瑞睿)\thanks{Correspondence: 1 Zhongziyuan Road, Dalang, Dongguan, Guangdong 523000, China. Email: fanrr@ihep.ac.cn}} 
\author[3]{W. Michael Snow}
\author[1,2,4]{Xin Tong (童欣)\thanks{Correspondence: 1 Zhongziyuan Road, Dalang, Dongguan, Guangdong 523000, China. Email: tongxin@ihep.ac.cn}}

\affil[1]{Institute of High Energy Physics, Chinese Academy of Sciences, 19B Yuquan Road, Shijingshan District, Beijing, 100049, China}
\affil[2]{China Spallation Neutron Source, 1 Zhongziyuan Road, Dalang, Dongguan, Guangdong, 523803, China}
\affil[3]{Indiana University, Bloomington, IN 47405, USA}
\affil[4]{Guangdong Provincial Key Laboratory of Extreme Conditions, Dongguan, 523803, China}
\affil[5]{State Key Laboratory of Particle Detection and Electronics, Beijing, 100049, China}

\date{} 

\begin{document}
\begin{CJK}{UTF8}{gbsn}

\maketitle
\vspace{-3em} 

\noindent\rule{\textwidth}{0.4pt}

\begin{abstract}
\vspace{-2em} 
Polarized eV neutrons can address interesting scientific questions in nuclear physics, particle physics, and astrophysics/cosmology. We present the first experiment to polarize the neutrons on the Back-n beamline at the Chinese Spallation Neutron Source (CSNS) using an \textit{in-situ} NSF using spin-exchange optical pumping (SEOP) of $^{3}$He. A $^3$He polarization of 68\%±0.7\% for this \textit{in-situ} NSF was measured through neutron transmission method at Back-n.This is high enough to enable new experiments on the Back-n beamline.
\\
\textbf{Polarized neutron, Polarized nuclei, CSNS, White neutron, Fundamental physics research, Parity Violation}
\\
\textbf{PACS number(s):}  24.80.+y, 67.30.ep, 29.27.Hj ,  24.70.+s , 32.10.Dk 
\end{abstract}

\begin{multicols}{2}

\section{Introduction} 
\setlength{\parindent}{0em}

The discovery of parity violation in the decay of polarized $^{60}$Co in the late 1950s shocked physicists and motivated a reexamination of the validity of discrete symmetries in physics which remains a vital research frontier. Soon after the discovery of parity violation, in 1964, indirect time reversal symmetry violation was discovered in the 2$\pi$ decay of the $K^0$ meson\cite{2piK} which was later accommodated into the CKM matrix. As shown by A. Sakharov\cite{Sakharov} in 1967, CP-violation was demonstrated to be an essential criterion required to produce the matter-antimatter asymmetry in the early stage of the universe in Big Bang theory. However the CP violation seen in $K^0$ decay was several orders of magnitude smaller than the size needed to explain the lack of anti-matter we see in today's universe. Thus, new sources of CP/T violation are required to explain the currently observed baryon asymmetry of the universe within the Big Bang theory. 

Starting in the 80s, scientists started the construction of the theory and the experimental search for parity violation in p-wave resonances in heavy nuclei. Very large P-odd effects were discovered in $^{81}$Br, $^{111}$Cd, $^{117}$Sn, $^{139}$La and $^{238}$U at Dubna~\cite{Alfimenkov1983} and at KEK. In the 90s, the TRIPLE collaboration performed a very broad survey of parity violation in heavy nuclei at LANSCE. They scanned through many nuclei with a neutron beam with energies of 1 eV to several keV, polarized using a polarized proton spin filter at their p-wave resonances. They discovered more than 75 P-odd asymmetries in p-wave resonances and were able to perform a statistical analysis of the distribution of results to show that the observed effects were consistent with the expected size of the nucleon-nucleon weak interaction\cite{PVcompnuc}. The ongoing work of the Neutron Optics Party and Time Reversal EXperiment (NOPTREX) collaboration builds upon the previous work at Dubna, KEK, and LANSCE to continue the search for PV in p-wave resonances and to perform TRIV searches. The Chinese NOPTREX collaboration at China Spallation Neutron Source (CSNS) aims to utilize the special properties of the Back-n white neutron source\cite{Back-n} and $^3$He neutron polarizer\cite{ChuyiNSF2021} to perform PV and TRIV experiments in China. 

Parity Violation (PV) and Time Reversal Invariance Violation (TRIV) can be probed using polarized neutrons in the p-wave resonances of nuclei. More detailed discussions on such new physics search exists\cite{BowmanGudkov2014, SnowHaddock2022}. Both PV and TRIV can be amplified in p-wave resonances by very large factors of $10^4 - 10^6$  compared to their size in NN interactions\cite{Alfimenkov1983,Bunakov&Gudkov1983,Gudkov1992}. The amplification occurs through the mixing of p-wave resonances (neutron orbital angular momentum: l=1) and s-wave resonances (l=0) through a parity-odd matrix element. This resonance amplification mechanism makes the compound nucleus a very sensitive system to probe fundamental interactions and is complementary to neutron or nuclear EDM searches\cite{EDMnewP,EDMBeyond,PSInEDM,USTCYbEDM} which also search for TRIV.

For a neutron of spin $\vec{\sigma_n}$, momentum $\vec{k_n}$ incident on a target nucleus of spin $\vec{I}$, we can construct an expression for the forward scattering amplitude $f$ as the following\cite{BowmanGudkov2014}:
\begin{align}
\label{Eqn:NNfoward}
f = &f_0 + f_1(\vec{\sigma_n}\cdot{}\vec{I}) + f_2(\vec{\sigma_n}\cdot\vec{k_n})\\
&+ f_3(\vec{\sigma_n}\cdot[\vec{k_n}\times\vec{I}]) + f_4(\vec{\sigma_n}\cdot(\vec{I}\times\vec{k_n})(\vec{I}\cdot\vec{k_n}))
\end{align}
Notice each term reacts differently under parity and/or time-reversal. $f_0$ is P-even/T-even and spin-independent, $f_1$ is P-even/T-even but spin-dependent (leading to the so-called \lq\lq pseudomagnetic precession \rq\rq), $f_2$ is P-odd/T-even the parity violating time-reversal conserving term, $f_3$ is P-odd/T-odd parity and time-reversal violating term and $f_4$ is P-even/T-odd parity conserving time-violating term. A more detailed discussion of the formalism shown above exists~\cite{Hiro&Gudkov2018, Hiro&Gudkov2020}.

The different symmetries of each term access different physics and require different experimental setups to observe. As we can see from Eqn.\ref{Eqn:NNfoward}, all the terms excluding the most trivial one $f_0$ have a neutron helicity ($\vec{\sigma_n}$) dependence which makes a neutron polarizer a critical and essential tool to probe such terms especially $f_3$ and $f_4$that may lead to a beyond standard model discovery. From thermal to epithermal neutron energies, the SEOP \textit{in-situ} $^3$He polarizer is the optimum choice for neutron polarization due to its high neutron polarization efficiency, low cost and convenience in installation relative to a proton polarizer that usually requires a high-power liquid helium cryostat to operate. 

The optical theorem of quantum scattering theory relates the imaginary part of the forward scattering amplitude to the total cross-section.
\begin{equation}
\sigma_{tot} = \frac{4\pi}{k}Im[f(0)]
\end{equation}
The forward scattering amplitude $f$ describes how initial state $\ket{i}$ and final state $\ket{f}$ are connected by the parity violating $V_P$ and parity and time reversal violating $V_{PT}$ weak interaction potentials: 
\begin{equation}
    f = \bra{f}V_P + V_{PT}\ket{i}
\end{equation}
Weak mixing matrix elements $v$ and  $w$ are obtained by acting s- and p- wave resonance wave functions $\ket{\phi_s}$ and $\ket{\phi_s}$ on   $V_P$ and $V_{PT}$:
\begin{equation}
    v+iw = \bra{\phi_p}V_P+V_{PT}\ket{\phi_s}
\end{equation}
The connection between terms $f_2$ and $f_3$ in the approximation where the ixing with the p-wave amplitude is dominated by one nearby s-wave amplitude is then illustrated by the following\cite{Hiro&Gudkov2018}:
\begin{equation}
    \frac{\Delta \sigma_{PT}}{\Delta \sigma_P}=\kappa(J)\frac{w}{v} 
\end{equation}
for which $\kappa(J)$ is an amplification term and $w$, $v$ are the weak mixing matrix elements. $\Delta \sigma_{PT}$ and $\Delta \sigma_P$ are the difference of $f_2$ and $f_3$ total cross-section by varying neutron helicity going into target nuclei which is directly proportional to the size of observed asymmetries. Term $f_2$ and $f_3$ are both amplified at p-wave resonances by the mixing of s- and p-wave resonances for $10^4$ to $10^6$ times. Although through different channel, the same mixing mechanism lets us connect the size of effect of these two types of asymmetries. By measuring $\Delta \sigma_P$ and $\kappa(J)$, we can estimate the size of $\Delta \sigma_{PT}$ for a particular p-wave resonance of a nuclei. Due to the statistical model of atomic nuclei, these quantities of interest along with resonance parameters are unique for each individual resonance. Hence, each p-wave resonance of interest should have their size of PV asymmetry $\Delta \sigma_P$ and amplification term $\kappa(J)$ measured in order to come up with a list of hierarchy for the future $\Delta \sigma_{PT}$ TRIV measurement. 

A helicty dependent resonance cross-section is defined as the following:
\begin{align}
\label{Eqn:sigma_pm}
    \sigma^\pm  = \sigma_p [1 \pm P_n f_n ]
\end{align}
$\sigma^\pm $ is the helicity dependent resonance cross-section for which $+$ is defined as neutron polarization along its direction of travel and $-$ as the opposite, $\sigma_p$ is the p-wave resonance cross-section of a nuclear target, $P_n$ is the polarization of the neutron beam and $f_n$ is the parity-odd longitudinal analyzing power which is an intrinsic property of a p-wave resonance.

Longitudinal asymmetry $A_L$ can be measured by transmission of neutrons with opposite helicities through a target defined in the following equation:
\begin{equation}
\label{Eqn:Asym}
A_L = \frac{N_+ - N_-}{N_+ + N_-}
\end{equation}
where $N_{\pm}$ is the helicity dependent neutron transmission yield for which $+$ and $-$ are defined as neutron polarization aligned and anti-aligned with its momentum vector. $N_\pm = N_0 \ exp(-n\sigma^\pm)$ for $N_0$ is the detector yield when the p-wave cross section $\sigma_p = 0$, $n$ the number density of the nuclear target in nuclei/cm$^2$. $A_L$ can be rewritten as\cite{PVcompnuc}:
\begin{align}
\label{Eqn:A_LApprox}
    A_L &= tanh(-P_n f_n n \sigma_p) \\
            &\approx -P_n f_n n \sigma_p
\end{align}
Any non-p-wave resonances or p-waves without an asymmetry will exhibit a near zero behavior on the $A_L$ calculations while the actual effect of interest should reveal itself with an $A_L$ value that stands out from the $A_L = 0$ in nearby neutron energies.

In this paper we describe the first operation of a polarized $^{3}$He Neutron Spin Filter (NSF) on the Back-n beamline at the CSNS. The polarized neutron beam created by this device on the Back-n beamline can enable many new types of experimental investigations on the different spectrometers and detector systems installed at this facility. We begin by reviewing the operation principles of polarized $^{3}$He NSF and present the transmission measurements that demonstrate the effective operation of the polarized $^{3}$He NSF on the Back-n beamline. We then describe the spectrometer systems on the Back-n beamline which we expect to be most well-suited to take advantage of this new polarized beam capability. Finally we describe one example of such a scientific application in the conclusion and discussion section: the search for parity violation in polarized neutron interactions in p-wave neutron-nucleus resonance reactions. 

\section{Polarized $^3$He as a NSF}
Polarized $^3$He NSF is a widely used instrument for neutron polarization. $^3$He NSF is very useful for material science and neutron scattering research due to its high efficiency in polarizing Angstrom-wavelength neutrons. The less well known application of $^3$He NSF is to polarize neutrons to probe fundamental nucleon nucleon (NN) interactions in the compound nucleus\cite{Coulter3He, YamamotoT2019, YamamotoT2020}. 
An off-situ $^3$He NSF may have a high $^3$He polarization (\~85\%) to start with\cite{Takuya3He2020}, the advantage of an \textit{in-situ} system lies in it's ability to constantly pump a $^3$He cell that is in the beam, maintaining a stable polarization for an experiment and is free from cell change.

\textit{\textbf{$^3$He neutron transmission formalism}}

For unpolarized $^3$He gas, the neutron transmission $T_0$ is given by the following equation:
\begin{align}
\label{Eqn:Transmission}
    T_0 &= T_e \ exp(- O)= T_e \ exp( - n_{\text{He}} \ \sigma_{0} \ l)  \\
        &= T_e \ exp(-0.0732 \  p\  l \ \lambda)
\end{align}
$T_e$ is neutron transmission through an empty glass cell with no $^3$He gas. $O = - n_{He} \ \sigma_{0} \ l$ is the neutron opacity factor.  $O$ is linearly proportional to $n_{He}$ the number density of $^3$He gas, $\sigma_{0} \approx 3000\ \lambda$\cite{3HeNSF1999} is the total cross section of unpolarized $^3$He in barn, and $l$ the length of  $^3$He gas neutrons have to travel through. For $^3$He gas, $O$ can be expressed into a more convenient form: $O = -0.0732\  p\ l \ \lambda$ for which $p$ is the pressure of $^3$He at 20°C in bar, $l$ in cm, and $\lambda$ is the neutron wavelength in Å.  
Polarized $^3$He neutron transmission $T_n$ is given by:
\begin{align}
\label{Eqn:Pol$^3$He}
    T_n &= T_e \ exp( - O) \ cosh(O\ P_{\text{He}}) \\
        &= T_0 \ cosh(O \ P_{\text{He}})
\end{align}
for which $P_{\text{He}}$ is the $^3$He polarization and let's define the neutron transmission yield  $T_n' = T_n/T_e = \text{exp}(-O)\text{cosh}(O
\ P_{\text{He}})$.

The ratio between polarized $^3$He neutron transmission and unpolarized $^3$He neutron transmission is then:
\begin{equation}
\label{Eqn:PolFit}
    \tfrac{T_n}{T_0} = cosh(O\ P_{\text{He}})
\end{equation}
Polarization of neutrons $P_n$ through a spin filter is,
\begin{equation}
\label{Eqn:P_n}
    P_n = tanh(O\ P_{He})=\sqrt{1-\tfrac{T_0^2}{T_n^2}}
\end{equation}
We can use the figure of merit $Q$, 
\begin{equation}
 \label{Eqn:FOM}   
    Q = T_n \ P_n^2
\end{equation}
to evaluate the analyzing power of certain spin filter system.
Using Eqn.\ref{Eqn:PolFit}, through the ratio of $T_n$ and $T_0$, we can obtain the absolute $^3$He polarization of our \textit{in-situ} NSF system and the resulting neutron transmission $T_n$ and polarization $P_n$.

\begin{figure}[H]
    \centering
    \includegraphics[width=0.95\linewidth]{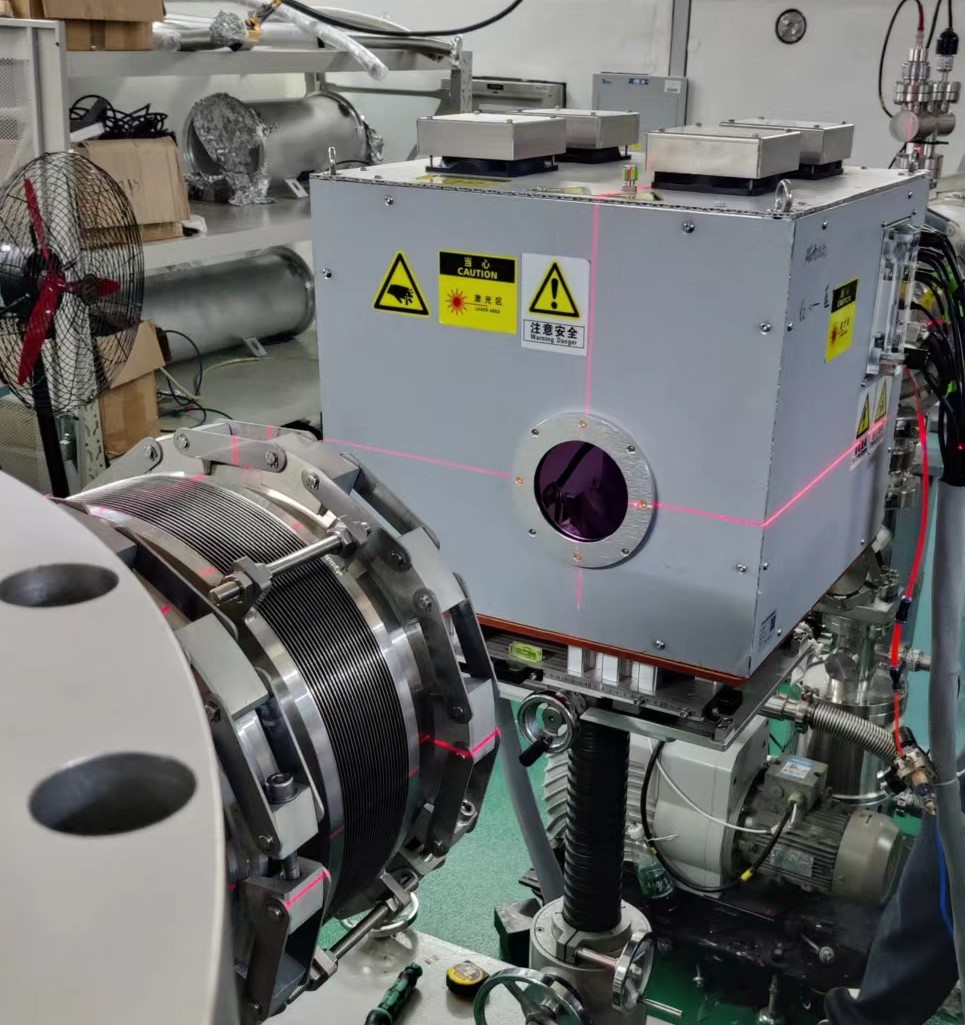} 
    \caption{\textit{in-situ} $^3$He SEOP NSF installed at ES\#1 of Back-n beamline} 
    \label{fig:3He}
\end{figure}

\textit{\textbf{$^3$He polarizer developed at CSNS}}
 CSNS neutron polarization device development started in 2019. Since then,  a spin-exchange optical pumping (SEOP)-platform\cite{ChuyiNSF2021}, a $^3$He glass cell workshop\cite{ZecongFilling2021} and further development in $^3$He polarization instrument techniques\cite{TLNSF2023} were conducted at CSNS. Currently, the group manages to produce a \textasciitilde8 cm long by $\phi$5 cm standard cylindrical $^3$He GE180 glass cell capable of holding 3.1 bar gas pressure at room temperature which can withstand 200 °C temperature. The previous \textit{in-situ} system achieves a 74\% $^3$He polarization as measured on BL-20 of CSNS\cite{Junpei2022}. The typical  $^3$He polarization fluctuation is $\sigma_{P_{He}} = \pm 2\%$ for the \textit{in-situ} system we developed at CSNS.

\begin{figure}[H]
    \centering
    \includegraphics[width=1\linewidth]{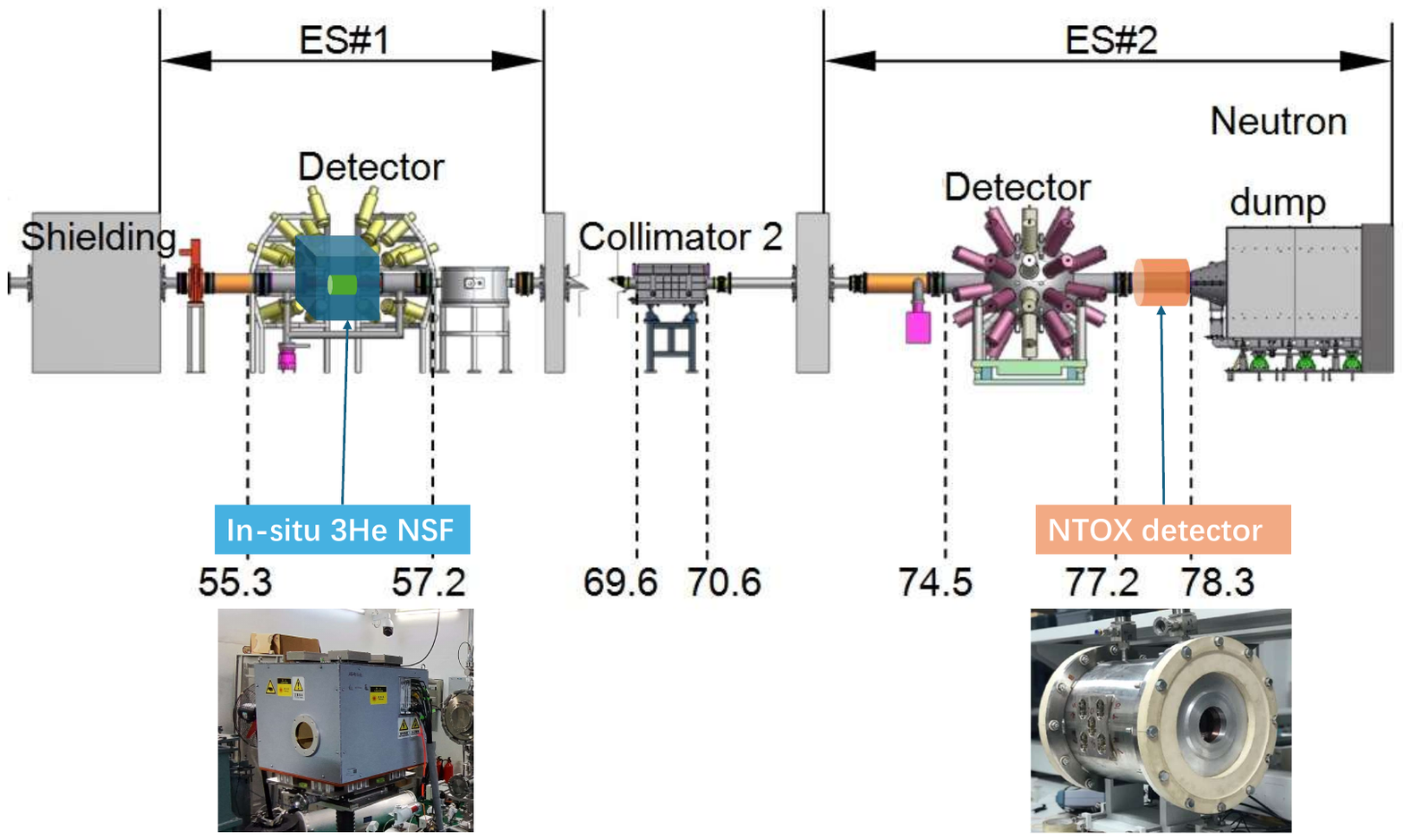} 
    \caption{This is the layout of this experiment. \textit{In-situ} $^3$He NSF was placed at ES\#1. NTOX detector was placed at ES\#2 right in front of beam dump.}
    \label{fig:layout} 
\end{figure}

\textit{\textbf{$^3$He transmission experimental results}}
In this experiment, we applied a new-designed \textit{in-situ} $^3$He NSF system developed at the CSNS on the Back-n White Neutron Source (WNS). This is the first time that an \textit{in-situ} $^3$He SEOP NSF system has been applied to the only WNS in China. From Fig: \ref{fig:layout} we can see where the $^3$He NSF and NTOX detector were placed. The $^3$He system was placed at a flight path of 55 m from the spallation target at ES\#1. NTOX installed at ES\#2 down stream of ES\#1 at 77.1 m flight path to measure neutron transmission through the $^3$He NSF. A cell with 2.53 bar pressure, 7.2 cm gas length was used in the \textit{in-situ} $^3$He SEOP system shown in Fig:\ref{fig:3He} \cite{Junpei2022}. The \textit{in-situ} $^3$He NSF system was deployed on the Back-n beam at location ES\#1 and the NTOX detector for the neutron flux measurement was located at ES\#2. The flight path distance of 77.1m from the spallation target was calibrated using a pre-established method\cite{NTOX}. 

Our neutron transmission experiment of the $^3$He \textit{in-situ} NSF system performed on Back-n beamline took 75 h of data, consisting of a polarized run of 57 h and an unpolarized run of 18 h. The $^3$He cell was pre-pumped before the experiment for in order to reach maximum polarization. The polarized $^3$He neutron transmission was measured first, with continuous pumping during the measurement to maintain the high $^{3}$He polarization. Second, depolarizing procedure was performed through period off-resonance Adiabatic Fast Passage (AFP) operation. Finally, the neutron transmission for the fully unpolarized in-situ $^3$He NSF was measured. 

A 1.7 mm thick BN (boron nitride) filter was applied upstream in the beamline to extend the neutron energy cutoff to below 19 meV, much lower than the \~0.3 eV cutoff energy than is normal on the Back-n beam, which usually employs a Cd+Ag+Co filter configuration. Frame overlap was observed for neutron energies lower than 19.4 meV, so data lower than this energy was ignored in our analysis. This was the first experiment of Back-n with such a filter configuration. Normalization of the neutron transmission data was performed using the number of protons injected to the spallation target. 

\begin{figure}[H]
    \centering
    \includegraphics[width=1\linewidth]{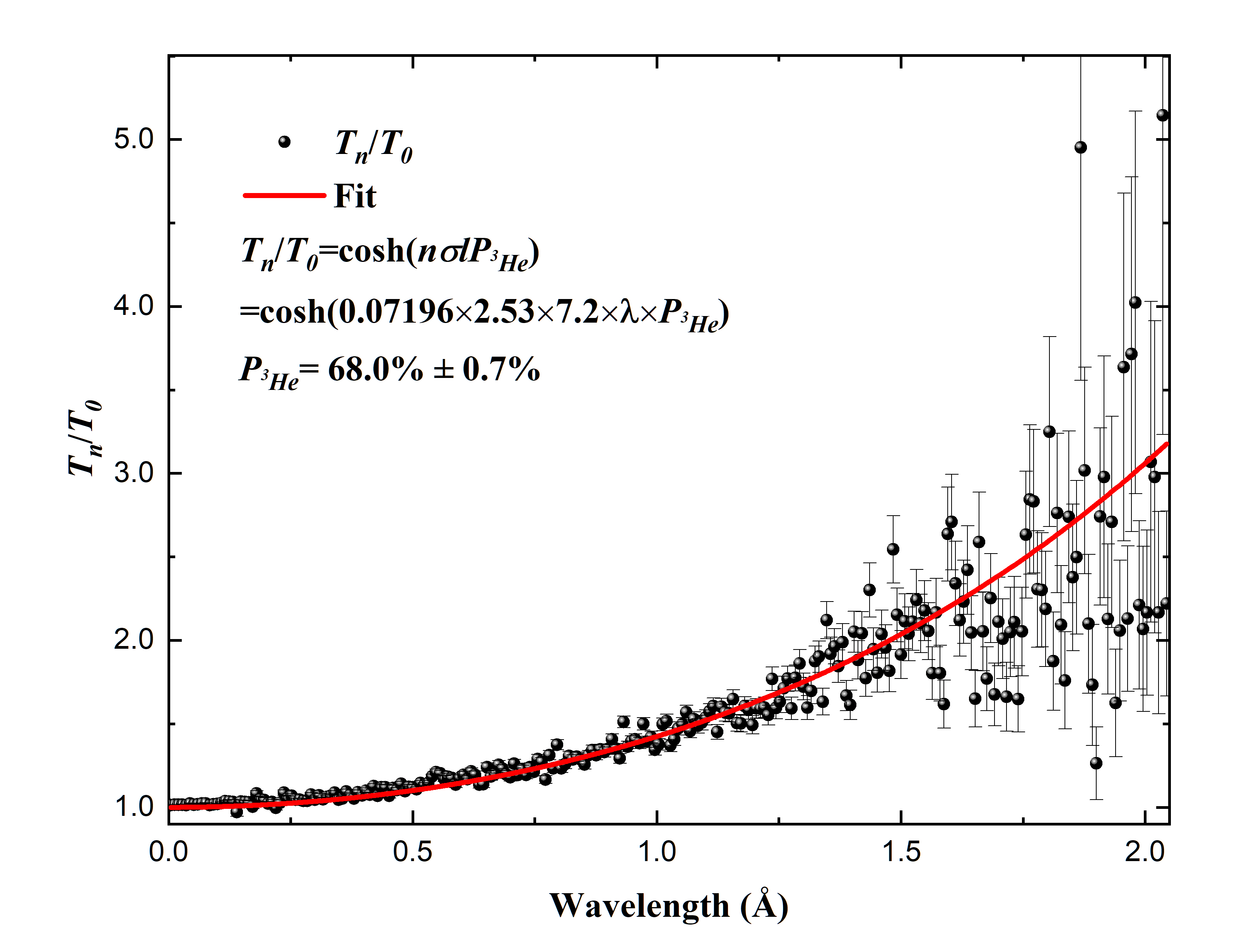}
    \caption{Fitting result for the $^3$He polarization measured on Back-n with the NTOX detector. The y-axis is the ratio of polarized($T_n$) vs unpolarized($T_0$) $^3$He transmission data. $T_n$ refers to the polarized $^3$He transmission. The x-axis is neutron energy in wavelength (Å). }
    \label{fig:results}
\end{figure}
The result shown in Fig:\ref{fig:results} is the plot of ratio of polarized $^3$He transmission data and unpolarized $^3$He transmission data fitted using Eqn.\ref{Eqn:PolFit}. All the data are normalized with the number of protons injected to the spallation target as discussed in the previous section. The error bar on $P_{^3He}$ is purely from fitting and does not include any systematic error. The larger error bars in $T_n/T_0$ at longer wavelengths in Fig:\ref{fig:results} comes from the significant decrease in neutron flux near the cutoff. The neutron flux of Back-n beamline from 0.3 to 0.9 Å (10$^{-1}$ - 10$^0$ eV) is about 10 to 10$^2$ lower than higher neutron energy regions \cite{Back-nFlux}.

\textbf{\textit{$^3$He system on Parity Violation measurement}}

Fig. \ref{fig:P,T,Q} shows the theoretical performance of a polarized $^3$He cell of \textit{p}=2.53 bar, \textit{l}=7.2  cm and $P_{\text{He}}$ = 68\% in the measured \textit{in-situ} NSF system. Figure of merit \textit{Q} peaks at 0.038 eV indicates this NSF balances polarization and transmission near this energy. 

The effect  of neutron polarization fluctuation $\sigma_{P_n}$ on the relative error of $A_L$ is:
\begin{align}
\label{Eqn:relA_L}
     (\frac{\sigma_{A_L}}{A_L})^2 \propto (\frac{\partial A_L}{\partial P_n} \frac{\sigma_{P_n}}{P_n})^2 
\end{align}
using the approximated form of $A_L$ from Eqn.\ref{Eqn:A_LApprox} we get:
\begin{align}
\label{}
\frac{\partial A_L}{\partial P_n} \frac{\sigma_{P_n}}{P_n} = -n\sigma_{p}f_n \frac{\sigma_{P_n}}{P_n}
\end{align}
using Eqn. \ref{Eqn:P_n} we can get the effect of $\sigma_{P_{He}}$ on the relative error of $P_n$:
\begin{align}
\label{relP_n}
(\frac{\sigma_{P_n}}{P_n})^2 \propto (\frac{\partial P_n}{\partial P_{He}} \frac{\sigma_{P_{He}}}{P_{He}})^2
\end{align}

\begin{figure}[H]
    \centering
    \includegraphics[width=1\linewidth]{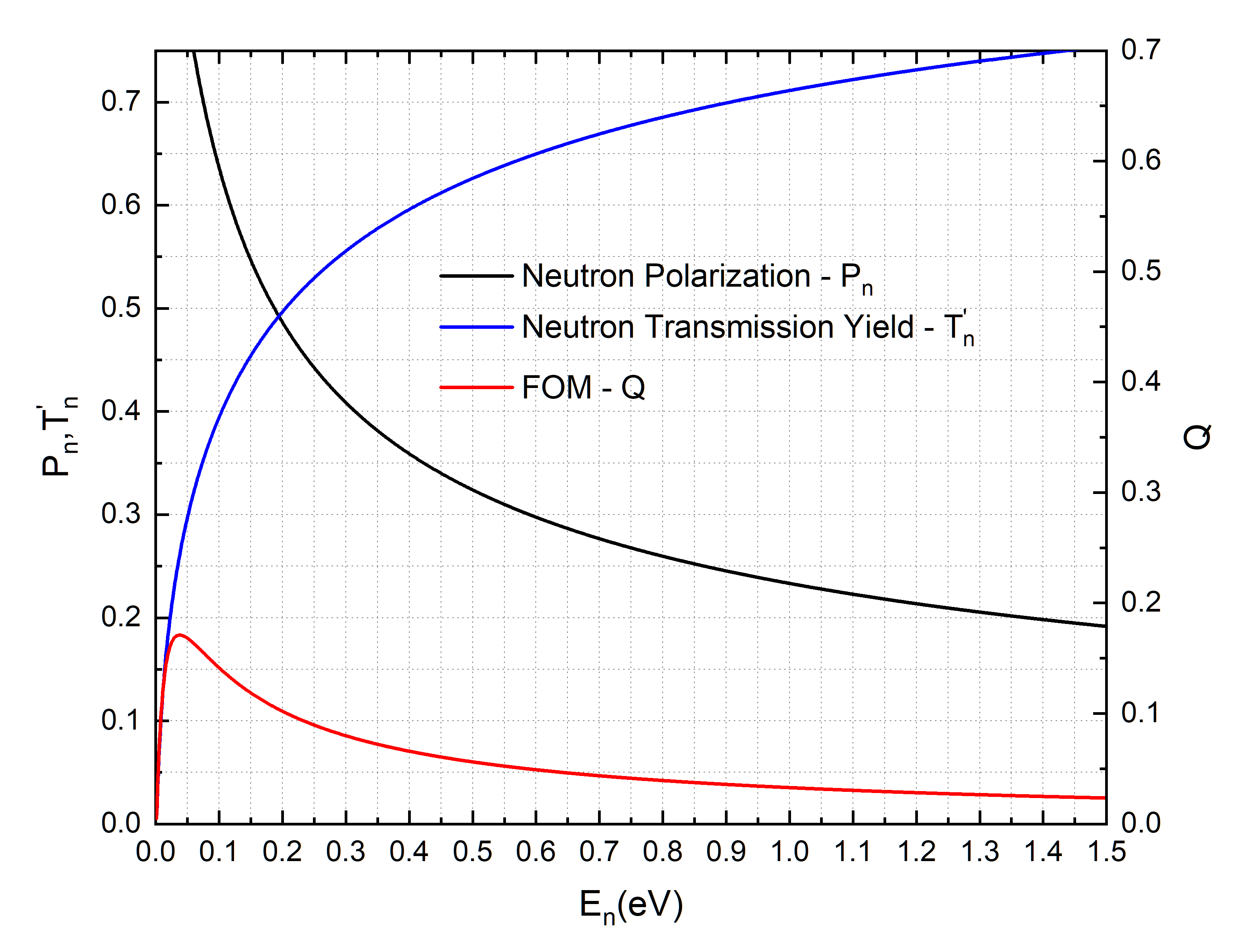}
    \caption{Theoretical performance of the measured $^3$He NSF system with: Neutron Transmission Yield $T_n'$ , Neutron Polarization $P_n$ and Figure Of Merit (FOM) $Q$ as a function of Neutron Energy $E_n$}
    \label{fig:P,T,Q}
\end{figure}

If we do an first order approximation on Eqn.\ref{Eqn:P_n} of $P_n$  just like what we did to to $A_L$. At thermal to epithermal energy, the contribution of neutron energy uncertainty is significantly suppressed so $O$ is treated as a constant here. We then get:
\begin{align}
\label{}
\frac{\partial P_n}{\partial P_{He}} \frac{\sigma_{P_{He}}}{P_{He}} = O = -0.0732pl\lambda
\end{align}
Combining equations above, we get the estimate of the effect of $^3$He polarization fluctuation on $A_L$
\begin{align}
     (\frac{\sigma_{A_L}}{A_L})^2 &\propto (0.0732*pln_{target}\sigma_{p}f_n\lambda \frac{\sigma_{P_{He}}}{P_{He}})^2 \\
     &\propto ( \frac{0.0209* pln_{target}\sigma_{p}f_n}{\sqrt{E_n}}    \frac{\sigma_{P_{He}}}{P_{He}})^2
\end{align}
Eventually, we can estimate the rough size of relative error from $^3$He polarization fluctuation contributing to the relative error of $A_L$. Using the \textit{in-situ} NSF system we discussed before for which: $p=2.53$ bar, $l=7.2$ cm, $\sigma_{P_{He}} = \pm 2\%$ and $P_{He}=68\%$, for a $^{nat}$La of 2cm thickness $n_{La} = 5.34 \times 10^{22}$ nuclei/cm$^2$,  $\sigma_p( 0.734eV)= 13.6$ barns, $f_n = 0.084$\cite{Coulter3He}, and $E_n = 0.734$ eV, we get the contribution to be $0.0059\%$. Hence, such fluctuation in $^3$He polarization will not be the major source of error in asymmetry measurements. 

A commonly used resonance in the measurement of PV as a check to a PV experimental setup is the 0.734 eV p-wave resonance of $^{139}$La since it has the largest measured parity violation asymmetry ($A_L \approx $ 10\%) at epithermal neutron region. From Fig. \ref{fig:P,T,Q} we can see at 0.734 eV there \~ 29\% of neutron polarization P. Previous NOPTREX experiments at J-PARC on BL-04\cite{Takuya3He2020} used an off-situ $^3$He cell that provides \~27\% neutron polarization at 0.734eV when at maximum polarization. This current \textit{in-situ} polarizer developed at CSNS is capable to carrying out PV measurements for near 1eV p-wave resonances. At 10 eV, only 8\% neutron polarization remains. We can still measure $A_L$ to a desired statistical accuracy but with the expense of greatly increase the required beam time. We can see from Eqn.\ref{Eqn:P_n} we can improve neutron polarization $P_n$ as either the neutron polarization $P_{He}$ or the opacity factor $O$ increases. The simplest solution is to increase $O$ by increasing neutron travel $l$ through polarized $^3$He gas which can be achieved through stacking two or more \textit{in-situ} systems in the beamline. 

\section{Back-n Beamline Properties and Instrumentation}
\begin{figure}[H]
    \centering
    \includegraphics[width=1\linewidth]{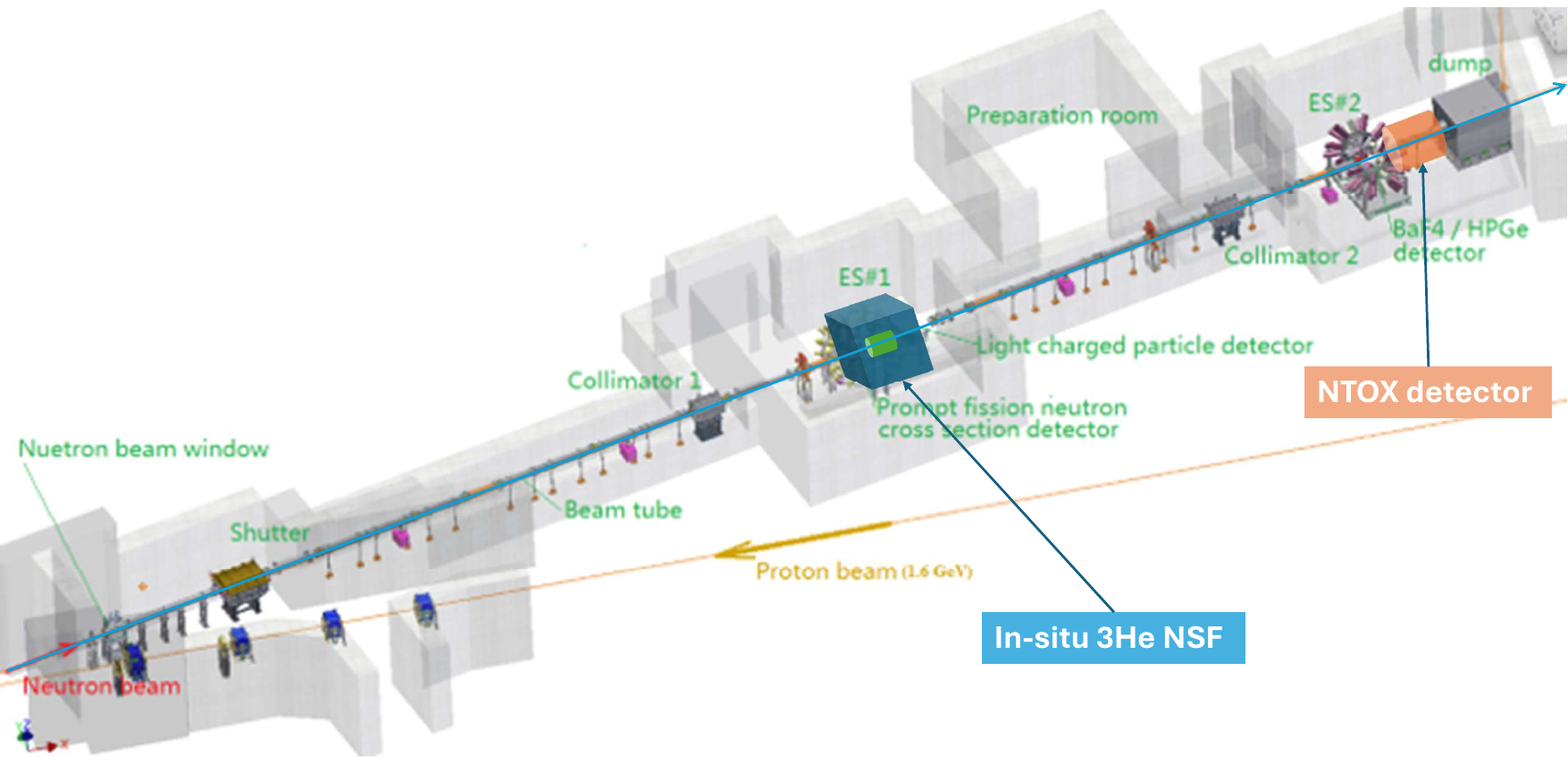} 
    \caption{Back-n beamline layout with indication of the location of $^3$He NSF and NTOX detector} 
    \label{fig:Back-n} 
\end{figure}
The CSNS Back-n WNS is an unique beamline due to its wide energy range of incident neutrons, from thermal to 300 MeV, and its long flight path: approximately 54 meters to ES \#1 and 77 meters to ES \#2. It also features a high neutron flux of $10^4 - 10^7 n/cm^2$ when operating at 125 kW power with a 25 Hz double bunch pulse \cite{Back-n}.  The NOPTREX collaboration consider the Back-n beamline most suitable for Parity and Time violating asymmetry measurements due to its high flux, long flight path, low beam divergence and spacious ES\#1 and ES\#2 which opens up many possible experimental setups.

\textit{\textbf{Neutron total cross-sectional spectrometer (NTOX)}}

\begin{figure}[H]
    \centering
    \includegraphics[width=1\linewidth]{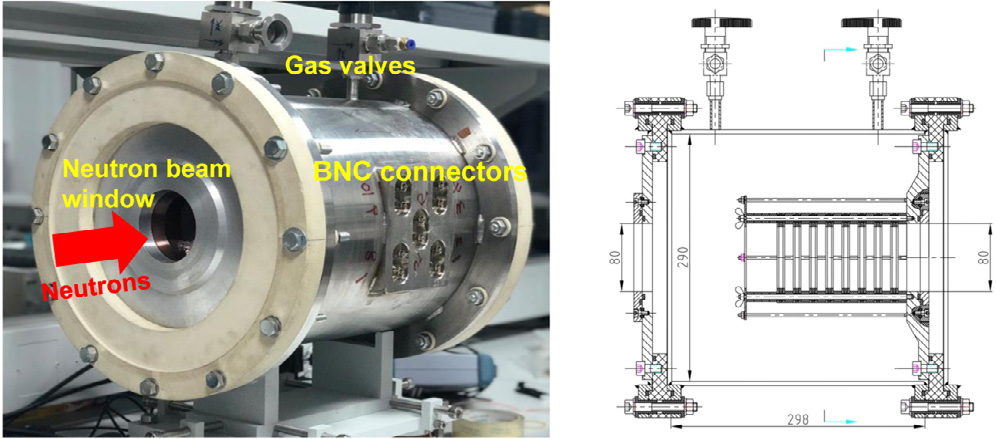} 
    \caption{NTOX neutron total cros-section spectrometer} 
    \label{fig:NTOX} 
\end{figure}
The NTOX detector\cite{NTOX} shown in Fig.\ref{fig:NTOX} is basically a multilayer fast fission chamber for fission cross-section or neutron total cross-section measurement usually setup at the ES\#2 location.
\\
\textit{\textbf{GTAF-II (n,$\gamma$) spectrometer}}
\begin{figure}[H]
    \centering
    \includegraphics[width=1\linewidth]{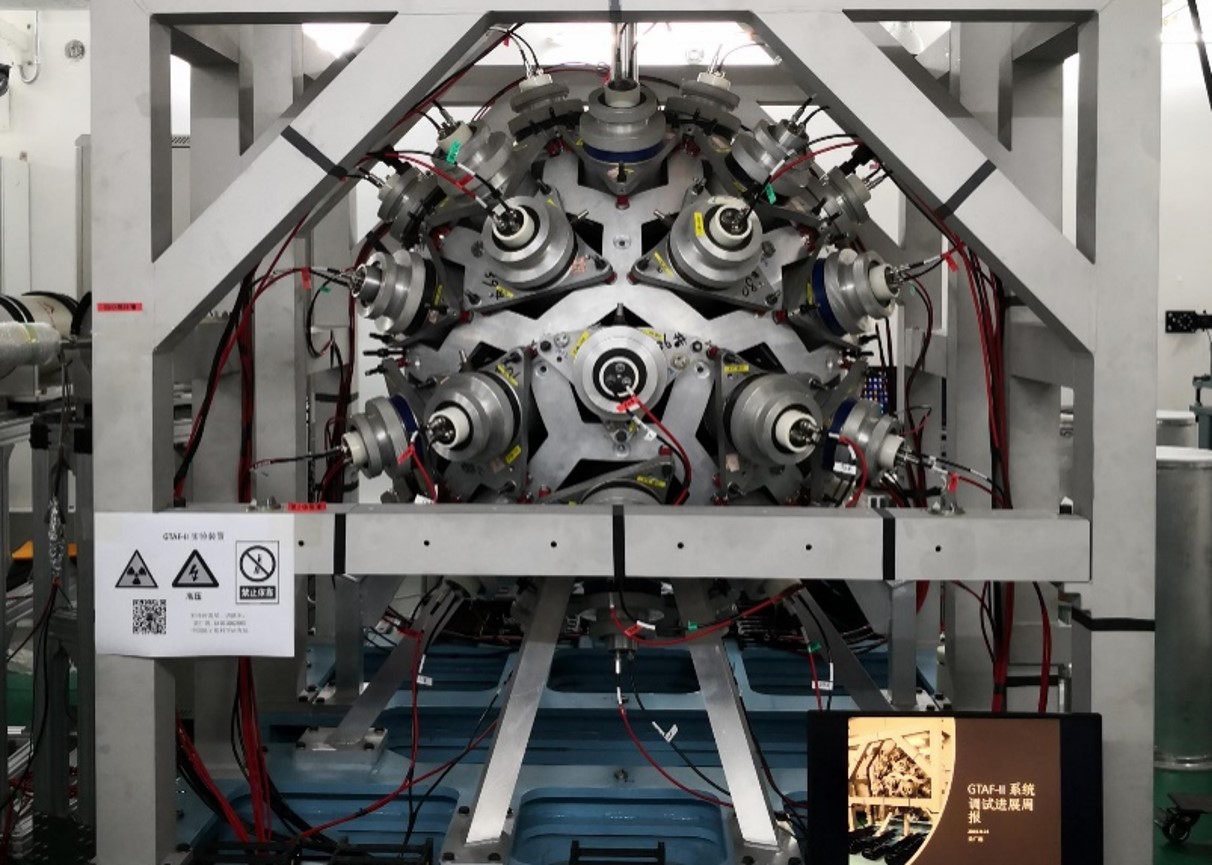} 
    \caption{GTAF-II 4$\pi$ BaF$_2$ (n,$\gamma$) spectrometer} 
    \label{fig:GTAF} 
\end{figure}
The Gamma Total Absorption Facility - II (GTAF-II) shown in Fig.\ref{fig:GTAF} is a 4$\pi$ BaF$_2$ (n,$\gamma$) spectrometer located at location ES\#2 of Back-n. It consists of 42 detector elements of which 40 are BaF$_2$ scintillation detectors capable of covering near 90\% of 4$\pi$ solid angle\cite{GTAF}. The near unity detection efficiency combined with the high intensity of WNS can perform efficient searches for neutron parity violation in p-wave resonances. Its high precision in neutron energy measurement and the capability of measuring the gamma angular distribution is invaluable in probing T-violating terms in spin-angular correlations between incoming neutrons and outgoing gamma-rays. The combination of a polarized eV neutron beam and a $4\pi$ BaF2 gamma calorimeter on an intense spallation neutron source like the CSNS is unique in the world and will enable new physics measurements.  

\section{Conclusion}
An \textit{in-situ} SEOP $^{3}$He neutron polarizer was successfully operated on the Back-n beamline and $^3$He neutron transmission under different conditions was measured. The $^3$He NSF achieved a $68\% \pm 0.7\%$ maximum polarization. This successful demonstration opens up various different types of future possible experiments probing fundamental symmetries at the Back-n WNS. 

\section{Discussion}

In this section, we present a possilbe future experiment on the Back-n beamline PV and TRIV experimental setups which only need the addition of a $^3$He NSF of the type that we have just described. A description of the extensive range of possibilities for combining a WNS with a neutron polarizer warrants a separate article. 
\\
\textit{\textbf{P-odd/T-even($f_2$) experiment}} 
\begin{figure}[H]
    \centering
    \includegraphics[width=1\linewidth]{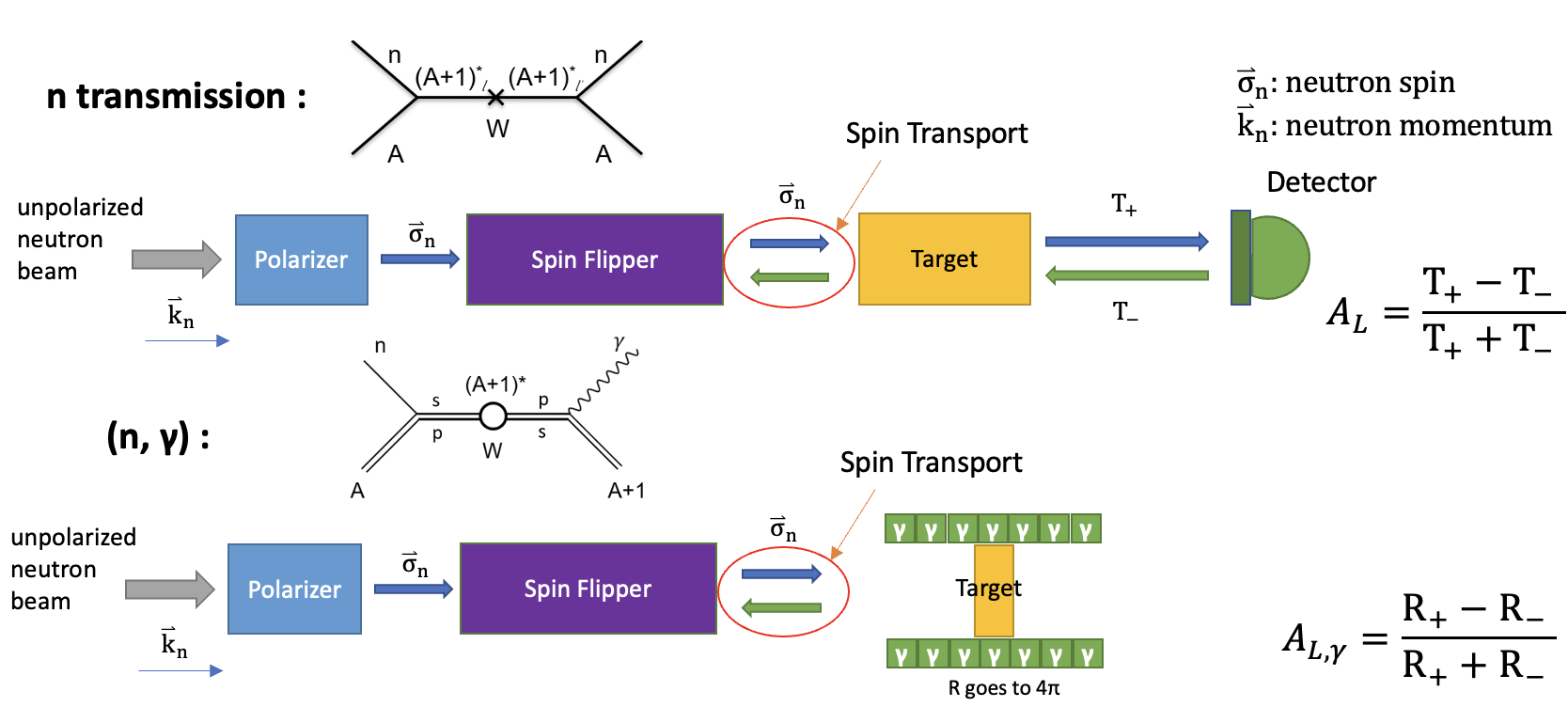} 
    \caption{PV experiment setups. There exists two methods shown in this figure to probe Parity Violation: the neutron transmission method\cite{PVcompnuc} and the $(n,\gamma)$ method\cite{NPDG}} 
    \label{fig:PVsetup} 
\end{figure}

In Fig. \ref{fig:PVsetup} is the setup which can probe the $f_2(\vec{\sigma_n}\cdot\vec{k_n})$ P-odd/T-even parity violating time reversal conserving term. By changing the direction of polarization of the $^3$He NSF, T$_{+/-}$ or R$_{+/-}$ can be measured and an asymmetry acquired as specified in Eqn:\ref{Eqn:Asym}. The two methods differ mainly in the means of detection from an instrumental point of view. At Back-n, the existing spectrometers that are most adequate for the requirements of both PV experiment methods are NTOX and GTAF shown in Fig.\ref{fig:NTOX} and Fig.\ref{fig:GTAF} respectively.

\section{Acknowledgement}
W.M.S. and M.Z. acknowledge support from the US National Science Foundation (NSF) grant PHY-2209481 and the Indiana University Center for Spacetime Symmetries. The work was supported by National Natural Science Foundation of China funded projects (12205326). This work was also supported by the National Key Research and Development Program of China (Grant No. 2020YFA0406000, No. 2020YFA0406004 and 2023YFA1606602). The $^3$He spin filter implemented in the experiment was developed within the Guangdong Provincial Key Laboratory of Extreme Conditions: 2023B1212010002 and the Dongguan Introduction Program of Leading Innovative and Entrepreneurial Talents (No. 20191122).

\end{multicols}
\end{CJK}

\end{document}